\documentclass[prb,amsmath,amssymb,reprint]{revtex4-1}

\usepackage{hyperref}
\usepackage{floatrow}
\usepackage{graphicx}
\usepackage{amsmath}
\usepackage{bm}
\usepackage{mhchem}

\begin{document}

\title{Impact of non-parabolic electronic band structure on the optical and transport properties of photovoltaic materials}

\date{\today}

\author{Lucy D. Whalley}
\affiliation{Department of Materials, Imperial College London, Exhibition Road, London SW7 2AZ, UK}

\author{Jarvist M. Frost}
\affiliation{Department of Physics, Kings College London, Strand, London WC2R 2LS, UK}

\author{Benjamin J. Morgan}
\affiliation{Department of Chemistry, University of Bath, Claverton Down, Bath BA2 7AY, UK}

\author{Aron Walsh}
\affiliation{Department of Materials, Imperial College London, Exhibition Road, London SW7 2AZ, UK}
\affiliation{Department of Materials Science and Engineering, Yonsei University, Seoul 03722, Korea}
\email{a.walsh@imperial.ac.uk}  

\begin{abstract}

The effective mass approximation (EMA) models the response to an external perturbation of an electron in a periodic potential as the response of a free electron with a renormalised mass. For semiconductors used in photovoltaic devices, the EMA allows calculation of important material properties from first-principles calculations, including optical properties (e.g.\ exciton binding energies), defect properties (e.g.\ donor and acceptor levels) and transport properties (e.g.\ polaron radii and carrier mobilities). The conduction and valence bands of semiconductors are commonly approximated as parabolic around their extrema, which gives a simple theoretical description, but ignores the complexity of real materials. 
In this work, we use density functional theory to assess the impact of band non-parabolicity on four common thin-film photovoltaic materials --- GaAs, CdTe, \ce{Cu2ZnSnS4} and \ce{CH3NH3PbI3} --- at temperatures and carrier densities relevant for real-world applications. First, we calculate the effective mass at the band edges. We compare finite-difference, unweighted least-squares and thermally weighted least-squares approaches. We find that the thermally weighted least-squares method reduces sensitivity to the choice of sampling density. Second, we employ a Kane quasi-linear dispersion to quantify the extent of non-parabolicity, and compare results from different electronic structure theories to consider the effect of spin-orbit coupling and electron exchange. Finally, we focus on the halide perovskite \ce{CH3NH3PbI3} as a model system to assess the impact of non-parabolicity on calculated electron transport and optical properties at high carrier concentrations. We find that at a concentration of $10^{20}\,\mathrm{cm}^{-3}$ the optical effective mass increases by a factor of two relative to the low carrier-concentration value, and the polaron mobility decreases by a factor of three. Our work suggests that similar adjustments should be made to the predicted optical and transport properties of other semiconductors with significant band non-parabolicity.

\end{abstract}

\maketitle

\section{Introduction}
Many semiconductor properties depend on the response of electrons to an external pertubation.
This perturbation could take the form of an electric field, change in temperature or an applied lattice stress.  
In a crystal, the response depends on the interaction of the electrons with a periodic potential.
The effective mass approximation assumes that this response of an electron in a periodic potential is equivalent to that of a free electron with a renormalised mass, which is called the ``effective mass''.
We emphasise that there is no direct relation between the actual mass of the electron and this effective mass, 
as the electron we refer to here is a quasi-particle composed of collective excitations of the interacting electrons. 
Many semiconductor device physics models, both semi-classical and fully quantum, are based on a band structure as specified by the effective mass parameter. 
Calculating the effective mass accurately from band structure calculations is therefore critical to correctly predicting optical and transport properties of semiconductors.
There has been renewed interest in this research area, focused in particular on the impact of electronic band structure anisotropy and non-parabolicity on the properties of thermoelectric materials.\cite{Gibbs2017,Mecholsky2014} Recent work has also emphasised the impact of band non-parabolicity on electron transport in semiconductors.\cite{Preissler2013,Krishnaswamy2017,Kang2018}

There are a number of algebraic definitions for the effective mass that can be used to calculate it from the band dispersion relation, $E(k)$; which can be obtained, for example, from \textit{ab initio} electronic structure calculations. For ideal parabolic bands, these definitions give equal effective mass values. For realistic non-parabolic bands, however, the calculated value of the effective mass depends both on the chosen definition and the numerical implementation (e.g.\ how $E(k)$ is discretised), which, in turn, changes the predicted values of optical and transport material properties. This issue is particularly pertinent for materials with applications that depend on optical or electronic performance, such as photovoltaics. Because of the role played by the effective mass in linking the fundamental electronic structure to key material properties, it is necessary to understand how the choice of definition and method for calculating the effective mass impacts first the calculated value and second any subsequently derived material properties. 

In this paper we discuss three definitions of effective mass: curvature, transport and optical.
The conventional definition of effective mass, which we will refer to as the curvature effective mass, $m_\text{c}$, is
\begin{equation} \label{curvature}
\frac{1}{m_\text{c}}= \frac{1}{\hbar^2}\frac{\partial^2E}{\partial k^2}.
\end{equation}

\begin{figure}[tb]
\includegraphics[width=\textwidth]{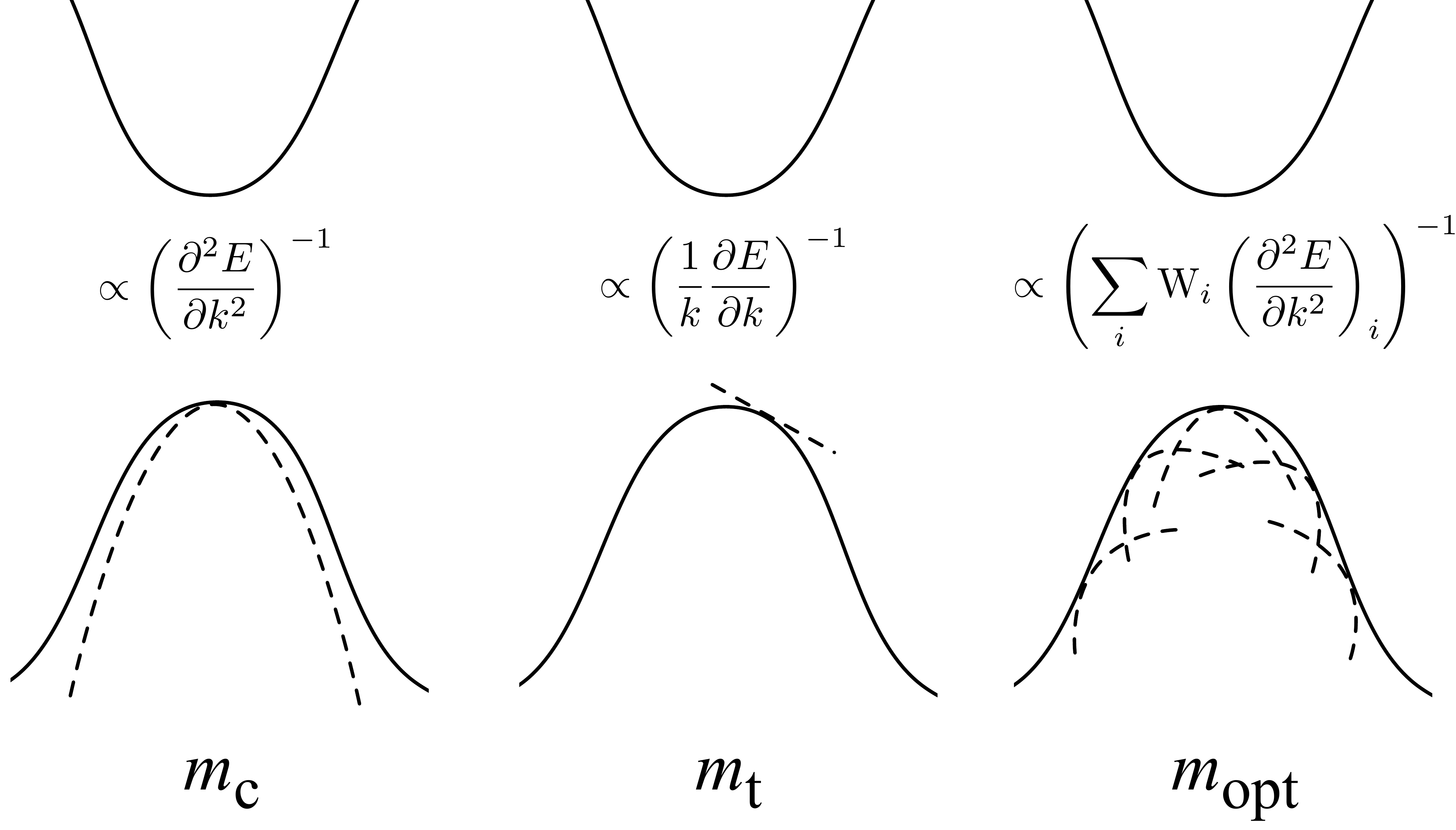}
\caption{\label{effmass_schematic} (left) the curvature effective mass $m_{\text{c}}$ is inversely proportional to the curvature of the electronic dispersion in reciprocal space; (middle) the transport effective mass $m_{\text{t}}$ is inversely proportional to the gradient of the electronic dispersion in reciprocal space; (right) the optical effective mass $m_{\text{opt}}$ is the average curvature effective mass, weighted according to the Fermi--Dirac distribution. }
\end{figure}

This expression is derived using Newton's second law\cite{Ashcroft1976,Ariel2012b} and so is commonly referred to as the inertial effective mass or conductivity effective mass because it describes the acceleration of an electron in an applied electric field.
Because $m_\text{c}$ is inversely proportional to the curvature of the electronic dispersion in reciprocal space (Fig.\ \ref{effmass_schematic}), 
this effective mass can, in principle, be calculated directly from \textit{ab-initio} band structures.

For semiconductors with low carrier concentrations we are often interested in the dispersion of eigenstates close to conduction or valence band extrema.
In these limits, the band dispersion is commonly approximated as parabolic, which gives an analytical expression of the curvature effective mass that is independent of sampling density:

\begin{equation} \label{parabolic}
E(k)= \frac{\hbar^2k^2}{2m_\text{c}}.
\end{equation}

Because true band dispersion relations are never exactly parabolic, the effective mass obtained from a particular band structure depends on the approach used to numerically evaluate Eqn.\ \ref{curvature}. 

In materials with high carrier concentrations, or at high temperatures, eigenstates far from the band extrema are accessed, and it becomes increasingly important for characterisations of the effective mass to incorporate non-parabolic effects. 
An expression for the effective mass of a non-parabolic band can be derived by considering the relationship between the momentum and velocity of an electron wavepacket, via $mv=\hbar k$:\cite{Ariel2012b}
\begin{equation} \label{transport}
\frac{1}{m_\text{t}} = \frac{1}{\hbar^2 k}\frac{\partial E}{\partial k},
\end{equation}
where $m_\text{t}$ is often referred to as the transport effective mass.
For parabolic dispersions, $m_\text{t}$ is equal to the curvature effective mass given in Eqn.\ \ref{curvature}. 

One approach to describing non-parabolic band dispersions is to keep the second order $k^2$ ellipsoidal energy surfaces and introduce a non-linear dependence on the energy, 
\begin{equation} \label{nonparabolic}
\frac{\hbar^2k^2}{2m_{\text{t},0}} = \gamma(E) = E + \alpha E^2 + \beta E^3+ \ldots,
\end{equation}
where $m_{\text{t},0}$ is the transport effective mass at the band edge.
Keeping only the first non-linear term gives the Kane quasi-linear dispersion relation\cite{Kane1957}

\begin{equation} \label{kane}
\frac{\hbar^2k^2}{2m_{\text{t},0}} = E(1 + \alpha E).
\end{equation}

\begin{figure}[tb]
\includegraphics[width=\textwidth]{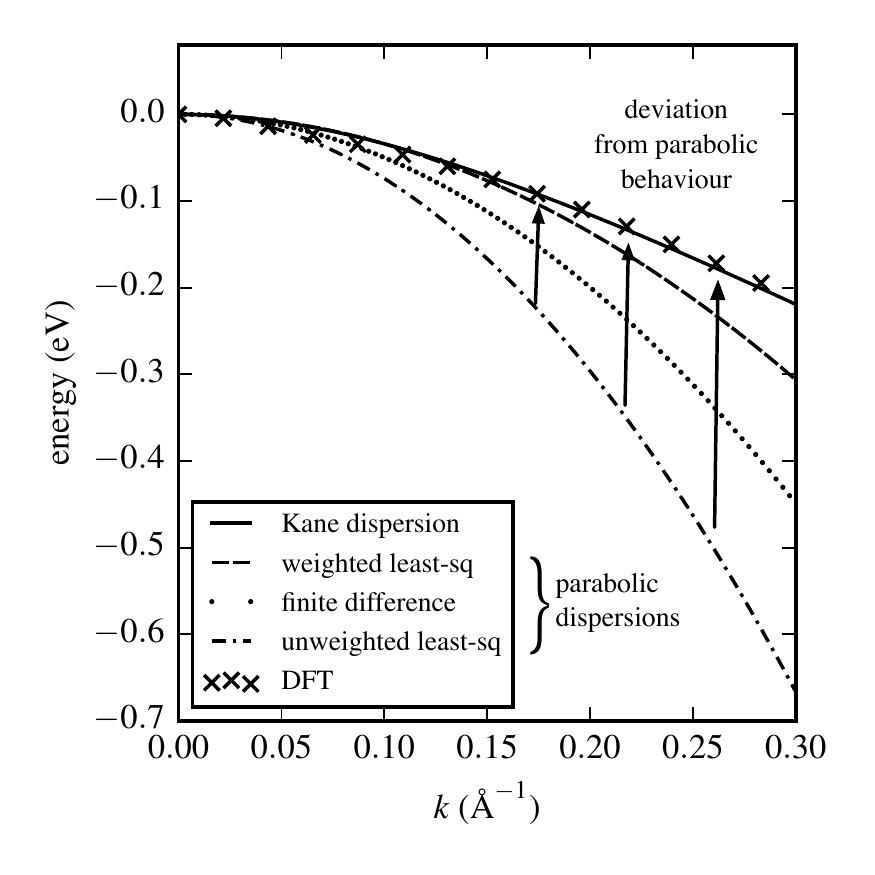}
\caption{\label{dispersion_fits} The electronic valence band energies in the [110] direction of \ce{Cu2ZnSnS4} and four approximate analytical representations of the dispersion relation. The parabolic dispersion relations depend upon the numerical method used (weighted least-squares fit---dash line, finite-difference---dot line and unweighted least-squares fit---dash-dot line). At higher binding energies the Kane quasi-linear dispersion (continuous line, Kane dispersion parameter $\alpha=2.2\,\mathrm{eV}^{-1}$) gives a better approximation to the DFT dispersion. Calculations use the hybrid exchange-correlation functional HSE06 with spin-orbit coupling.}
\end{figure}

The $\alpha$ parameter quantifies the amount of non-parabolicity due to flattening of the band as the eigenstate energy deviates from the band edge energy (Fig.\ \ref{dispersion_fits}). For a perfectly parabolic band, $\alpha=0$, while for a real conduction band $\alpha > 0$, and for a real valence band holes $\alpha < 0$. The band flattening corresponds to an increase in the transport effective mass. Differentiating Eqn.\ \ref{kane} with respect to $k$ gives the transport effective mass as

\begin{equation} \label{kanemass}
m_\text{t}(E) = m_{\text{t},0}(1+2 \alpha E).
\end{equation}

A third definition of effective mass is the optical effective mass, $m_\mathrm{opt}$, given by
\begin{equation}
\frac{1}{m_{\mathrm{opt}}} = \frac{2}{n_{\mathrm{e}}}\sum_{{l}}\sum_{{k}}^{\mathrm{{occ.}}} \frac{1}{m_\mathrm{c}^{{l}}(k)},
\end{equation}
where $n_{\text{e}}$ is the carrier concentration and $m_{\text{c}}^{{l}}$ is the curvature effective mass for band $l$ and occupied eigenstate $k$. 
Because the summation over all occupied eigenstates $k$ of each band $l$ accounts for any band non-parabolicity,
this definition has been used in the context of thermoelectric materials\cite{Gibbs2017} and transparent conducting oxides.\cite{Hautier2013} 
The summation over occupied states can be replaced by an integral along one-dimensional paths through reciprocal space, following a derivation by Huy \textit{et al}.,\cite{Huy2011}
\begin{equation} \label{opt2}
\frac{1}{m_{\mathrm{opt}}} = \frac{\displaystyle\sum_{{l}} \displaystyle\int f(E,T) \frac{\partial^2 E}{\partial k^2} dk}{\displaystyle\sum_{{l}} \displaystyle\int f(E,T) dk},
\end{equation}
where $f(E,T)$ is the Fermi--Dirac distribution for an eigenstate with energy $E$, in a system of particles with a Fermi level $E_{\mathrm{F}}$ and at temperature $T$,
\begin{equation} \label{fermidirac}
f(E,T) = \frac{1}{\exp\left(\frac{E-E_{\mathrm{F}}}{k_{\mathrm{B}}T}\right)+1}.
\end{equation}
In the case of only one occupied branch at $T=0\,\mathrm{K}$, we recover the transport effective mass evaluated at the Fermi level:
\begin{equation} \label{opt}   
\frac{1}{m_{\mathrm{opt}}} = \cfrac{\displaystyle\int_0^{k} \cfrac{1}{\hbar^2}\frac{\partial^2E}{\partial k^2} dk}{\displaystyle\int_0^{k}dk} = \frac{1}{\hbar^2k_{\mathrm{F}}} \frac{\partial E}{\partial k}\biggr\rvert_{k=k_{\mathrm{F}}},
\end{equation}
where $k_{\text{F}}$ is the Fermi wavevector.

When using the Kane dispersion as a more accurate approximation to the real material dispersion, the calculated transport and optical effective masses depend on the charge carrier energy.
As a result, when bands are progressively filled with charge carriers, through doping, photo-excitation, or increasing temperature, the corresponding effective mass will increase.\cite{Riffe2002}
Material properties that depend on the effective mass, such as electron and hole mobility, will also change as the carrier concentration is increased.
This contrasts with the behaviour under the parabolic approximation, where the effective mass is constant and independent of carrier concentration.
The extent to which the optical or transport effective masses increase as the carrier concentration increases depends on the magnitude of band non-parabolicity and the level to which the bands are filled. 
Band-filling can be measured experimentally via the Burstein--Moss band-gap shift.\cite{Burstein1954,Moss1954} 
This is a carrier-concentration--dependent shift in the optical absorption edge, which has been used to characterise \ce{GaAs} and other small--band-gap semiconductors since the 1970's. 

The Burstein--Moss band-gap shift can be measured using transient absortion spectroscopy, from which the effective mass and shape (parabolicity) of the electronic bands can be inferred.
Three experimental studies have measured the Burstein--Moss shift in the hybrid halide perovskite \ce{CH3NH3PbI3} (MAPI).\cite{Manser2014,Yang2015,Price2015} The resulting effective mass values range between $0.14\,m_{\text{e}}$--$0.30\,m_{\text{e}}$, where $m_{\text{e}}$ is the electron rest-mass.
Although each model differs in detail (Yang \textit{et al}.\ consider band-gap renormalisation,\cite{Yang2015} whilst Price \textit{et al}.\ consider photoinduced refractive index changes\cite{Price2015}), none of the models explicitly account for band structure non-parabolicity.
A previous computational study however has reported significant non-parabolicity in MAPI within an energy range accessible at room temperature ($k_{\mathrm{B}}T=26\,\mathrm{meV}$ at $T=300\,\mathrm{K}$).\cite{Brivio2014}

As we have discussed above, for any dispersion that is not parabolic, the calculated effective mass depends on the numerical method used to calculate band curvature and carrier concentration.
For materials with a significant number of excited carriers, the discrepancy between the parabolic and non-parabolic effective mass can be large.\cite{Ruf1990,Riffe2002}
An \textit{ab-initio}--derived effective mass that accounts for the non-parabolicity of a dispersion relation allows for a more accurate prediction of important material properties such as carrier mobility. 

As an example of how non-parabolicity can affect the theoretical description of semiconductor properties, we have considered four prototypical photovoltaic materials: \ce{CdTe}, \ce{GaAs}, \ce{CH3NH3PbI3} (MAPI) and \ce{Cu2ZnSnS4} (CZTS). 
For each material we calculate the curvature effective mass and Kane dispersion $\alpha$ parameter.
We compare calculations of the curvature effective mass using three different methods.
Two of these are widely used to calculate the curvature effective mass: a three-point finite-difference and a least-squares quadratic fit. 
The third method is a new approach which uses a least-squares quadratic fit weighted according to Fermi--Dirac statistics. 
We compare our results obtained from band structures calculated at different levels of theory, to assess the impact of spin-orbit coupling and choice of exchange-correlation treatment on the calculated values for effective mass and $\alpha$. 
We then focus on MAPI, and investigate the effect of non-parabolicity on the optical and transport properties at high carrier concentrations. 
We show that the Burstein--Moss shift is severely overestimated if calculated within the parabolic approximation, and that the Kane quasi-linear dispersion leads to significantly more accurate predictions.
We also calculate the optical effective mass and electron mobility over a carrier concentration range of $10^{16}$--$10^{20}\,\mathrm{cm}^{-3}$, which is the relevant range for concentrated solar power systems ($10^{17}$--$10^{18}\,\mathrm{cm}^{-3}$)\cite{Lin2018,Wang2018,Law2014}, or when excited under a laser for transient absorption or photoluminescence studies ($\sim10^{19}\,\mathrm{cm}^{-3}$).\cite{Richter2016}
We find that non-parabolicity leads to a significant change in transport properties for carrier concentrations above $10^{18}\,\mathrm{cm}^{-3}$.

\section{Methods}

\subsection{Electronic band structure calculations}

For each of the materials, \ce{CdTe}, \ce{GaAs}, \ce{MAPI} and \ce{CZTS}, optimised geometries and band dispersions were calculated using density functional theory (DFT) as implemented in the Vienna \textit{ab-initio} Simulation Package (\textsc{VASP}).\cite{Kresse1996} Valence wavefunctions were expanded in a plane-wave basis set with a cut-off of $500\,\mathrm{eV}$. Scalar-relativistic corrections for the core electrons were used within the projector augmented wave formalism.\cite{Blochl1994} The Brillouin zone was sampled using a Monkhorst-Pack $\Gamma$-centred $k$-point mesh. For CdTe, GaAs and MAPI a $6\!\times\!6\!\times\!6$ grid was used. For CZTS, which has a tetragonal crystal structure, a $6\!\times\!6\!\times\!4$ grid was used. 

Our initial structures were determined as follows: The \ce{GaAs} structure was taken from the Madelung handbook,\cite{Madelung2004}  \ce{CdTe}\cite{Rabadanov2001} and \ce{CZTS}\cite{Lafond2014} (in the tetragonal phase) were taken from X-ray diffraction data, whilst \ce{MAPI} was optimised starting from a pseudo-cubic (high temperature) structure available online.\cite{WMD} For all atomic relaxations a quasi-Newtonian algorithm and the PBEsol exchange-correlation functional was used. A different choice of exchange-correlation functional would result in distinct lattice parameters, which in turn, would affect the band dispersion. We have chosen to use the PBEsol exchange-correlation functional as it has been shown to accurately reproduce the experimental lattice parameters.\cite{Brivio2015}

For the band dispersions, a self-consistent electronic relaxation was followed by a non-self-consistent calculation along high symmetry lines,\cite{Setyawan2010} with points spaced $0.005\,\text{\AA}^{-1}$ apart in reciprocal space, except in the case of the hybrid HSE06 functional with spin-orbit coupling, where a spacing of $0.02\,\text{\AA}^{-1}$ was used. The total energy of each material was converged to within $10^{-6}\,\mathrm{eV}$. Exchange and correlation was modelled using: (i) the local density approximation (LDA); (ii) the PBEsol\cite{Perdew2008} generalized gradient approximation, and (iii) the screened hybrid functional HSE06.\cite{Heyd2003} Spin-orbit coupling (SoC) at the PBEsol and HSE06 levels of theory was introduced to investigate relativistic effects. Input and output files for the band structure calculations are available in two online repositories.\cite{Whalley2018,Whalley2018b}

\subsection{Calculation procedures for effective mass and polaron mobility}

For our calculations of effective mass and polaron mobiity, we considered one dimensional slices through the Brillouin zone; the effective mass tensor is reduced to a function of one variable $k$. Unless otherwise stated, the Fermi--Dirac distribution was calculated at $T=300\,\mathrm{K}$ with the Fermi level fixed half-way between the conduction band minimum and valence band maximum.  

The curvature effective masses were calculated using three approaches: i) a three point forward finite-difference method; ii) an unweighted quadratic least-squares fit\cite{Vanderwalt2011} to three points; iii) a quadratic least-squares fit, weighted according to the Fermi--Dirac distribution across all points up to energy $10\,k_{\mathrm{B}}T$ ($=0.26\,\mathrm{eV}$, at $300\,\mathrm{K}$). A mathematical expression for each approach is given in the Supplementary Information. 

To calculate the Kane dispersion parameters a sixth-order polynomial was fitted to the DFT eigenvalues over an energy range of $0.25\,\mathrm{eV}$ from the band edge. The first derivative of this continuous function was used to determine the transport effective mass. The transport effective mass was plotted against energy to give values for $\alpha$ and the effective mass at the band edge. The dispersion was truncated where the second derivative changes sign as this corresponds to an inflection point where the Kane dispersion is no longer valid. 

To calculate optical effective masses for MAPI, the integrand in Eqn.\ \ref{opt} was integrated with $E(k)$ given by the Kane dispersion. The exponential energy dependence in the Fermi--Dirac distribution means this integral converges quickly with $\Delta E(k)$. All effective mass calculations have been performed using the \textsc{effmass} package,\cite{Whalley2018a,Whalley2018b} 
which also includes a Jupyter notebook outlining the key calculation steps used in this paper.

Polaron mobilities for MAPI as a function of effective mass were calculated with a finite-temperature variational method, based on the Feynman path integral solution to the polaron problem. These were performed with the \textsc{PolaronMobility.jl} package.\cite{frost2017calculating,Frost2018,PolaronMobilityGithub}

\section{Results}

\subsection{Comparison of methods used to calculate effective mass at the band edge}

\newcommand{\ra}[1]{\renewcommand{\arraystretch}{#1}}
\begin{table*}[tb]\centering
\ra{1.3}
\begin{tabular}{@{}lclllclllclllclll@{}}\toprule
& \phantom{abc}&\multicolumn{3}{c}{finite-diff $m_\text{c}\left(m_{\text{e}}\right)$} & \phantom{abc}& \multicolumn{3}{c}{unweighted $m_\text{c}\left(m_{\text{e}}\right)$} &\phantom{abc} & \multicolumn{3}{c}{weighted $m_\text{c}\left(m_{\text{e}}\right)$} & \phantom{abc} & \multicolumn{3}{c}{$\alpha \left(\mathrm{eV}^{-1}\right)$}\\
\cline{3-5} \cline{7-9} \cline{11-13} \cline{15-17}
&& [100] & [110] & [111]$^*$  && [100] & [110] & [111]$^*$   && [100] & [110] & [111]$^*$   && [100] & [110] & [111]$^*$  \\ \colrule
GaAs\\
light hole 
&& 0.09 & 0.08 & 0.05        &&  0.08    & 0.07  & 0.07     && 0.09 & 0.08 & 0.08     && 3.32 & 3.64 & 3.69  \\
heavy hole 
&& 0.37 & 0.73 & 0.79        &&  0.31    & 0.58  & 0.25     && 0.32 & 0.75 & 0.53     && 0.38 & 3.37 & 1.70  \\
electron 
&& 0.07 & 0.07 & 0.06        &&  0.06    & 0.06  & 0.06     && 0.06 & 0.07 & 0.07     && 1.05 & 1.15 & 1.23  \\
CdTe\\
light hole 
&& 0.11 & 0.10 & 0.10        &&  0.11    & 0.10  & 0.10     && 0.12 & 0.10 & 0.10     && 1.25 & 1.49 & 1.64  \\
heavy hole 
&& 0.45 & 0.86 & 1.09        &&  0.44    & 0.83  & 1.20     && 0.45 & 0.83 & 1.06     && 0.38 & 0.99 & 0.77  \\
electron 
&& 0.09 & 0.09 & 0.09        &&  0.09    & 0.09  & 0.09     && 0.09 & 0.10 & 0.10     && 0.72 & 0.94 & 1.02  \\
MAPI\\
hole 
&& 0.28 & 0.15 & 0.14        &&  0.15    & 0.09  & 0.12     && 0.23 & 0.10 & 0.12     && 4.27 & 1.88 & 1.32  \\
electron 
&& 0.15 & 0.13 & 0.12        &&  0.18    & 0.10  & 0.19     && 0.19 & 0.10 & 0.18     && 2.21 & 1.35 & 0.16  \\
CZTS\\
hole 
&& 0.23 & 0.74 & 0.75        &&  0.22    & 0.54  & 0.60     && 0.32 & 1.16 & 1.23     && 3.96 & 2.21 & 1.50  \\
electron 
&& 0.19 & 0.19 & 0.19        &&  0.18    & 0.18  & 0.18     && 0.19 & 0.19 & 0.19     && 0.91 & 1.19 & 0.87  \\
\botrule
\end{tabular}

\caption{\label{largetable} The curvature effective mass at the conduction and valence band edges, calculated using three different methods as outlined in Methods, and $\alpha$ (a measure of band non-parabolicity). Calculations use the hybrid exchange-correlation functional HSE06 with spin-orbit coupling and a $k$-point spacing of $0.02\,\text{\AA}^{-1}$. The effective mass is calculated for the bands which are degenerate at the conduction band minima / valence band maxima; the corresponding band structures can be found in the SI. Spin-orbit coupling lifts the energy spin degeneracy and, in the case of MAPI, also leads to a Rashba splitting in reciprocal space. We take the mean average of the effective mass calculated for each split band. $^*$Direction is [001] in the case of the tetragonal crystal \ce{Cu2ZnSnS4}.}
\end{table*}

Two widely used methods for calculating the curvature effective mass are
the finite-difference approach, where the differential equation in Eqn.\ \ref{curvature} is approximated by a difference equation,
and the least-squares approach, where an unweighted second order polynomial is fitted to the DFT-calculated dispersion. 
For parabolic bands with negligible numerical noise, fitting to three points using a finite-difference or least-squares method gives an exact effective mass.
Eigenvalue energies obtained from DFT calculations, however, are subject to some degree of noise, due to numerical imprecision of numerical linear algebra methods. These errors are amplified by finite difference methods.\cite{Nearing2010} 
A non regularised least-squares approach also amplifies noise due to the $l_2$-norm used. 

For non-parabolic dispersion relations, curvature effective masses calculated using either finite-difference or least-squares methods depend on the sampling range and density in reciprocal space. 
We take the valence band edge of CZTS in the [100] direction as an example, calculated using PBEsol with SoC, and calculate the curvature effective mass, with the spacing in reciprocal space varied from $0.005\,\text{\AA}^{-1}$ to $0.025\,\text{\AA}^{-1}$.
Using an unweighted least squares fit to three points, the curvature effective mass varies from $0.05\,m_\text{e}$ to $0.08\,m_\text{e}$.
Using a three-point finite-difference method, the curvature effective mass varies from $0.06\,m_\text{e}$ to $0.12\,m_\text{e}$.

To reduce this sensitivity we propose using a least-squares fit with each point thermally weighted according to the Fermi--Dirac distribution.
This regularisation decreases sensitivity to the sampling density; 
for the previous example, the curvature effective mass is $0.07\,m_\text{e}$ across the whole $k$-point spacing range. 
The physically motivated weighting models the thermal distribution of electrons and attenuates the contribution far from the extrema, where changes in numeric sampling otherwise have a large effect on the calculated curvature effective mass. 
It provides a physically-intuitive energy range to fit over. 

Our calculated values for the curvature effective mass using the finite-difference and both least-squares fitting methods are listed in Table\ \ref{largetable}.
Conduction band electrons have light, highly isotropic masses and there is good agreement between all three approaches.
There are larger discrepancies between the effective mass values of valence band holes and we attribute this to the larger non-parabolicity of the valence band (see the $\alpha$ values in Table\ \ref{largetable}).

\begin{figure}[tb]
\includegraphics{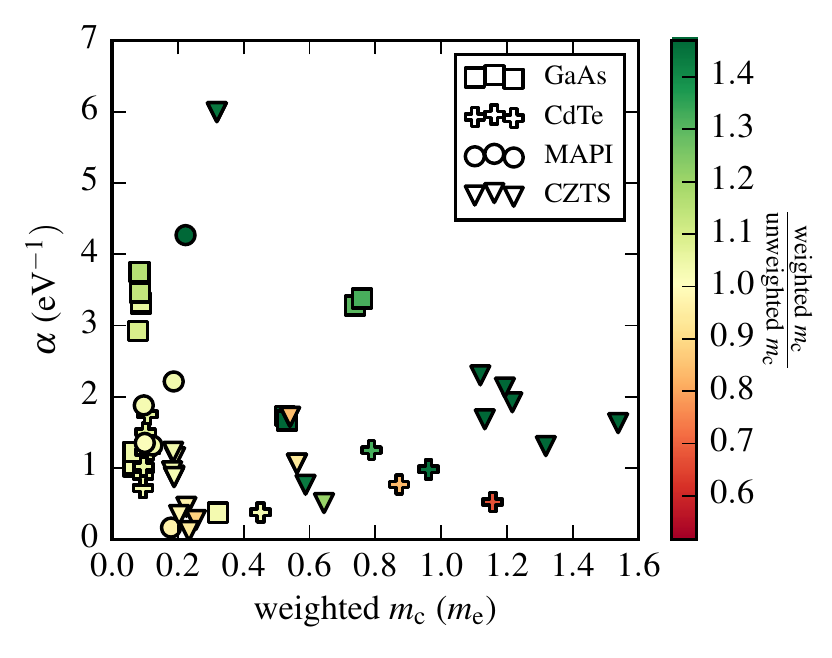}
\caption{\label{mass_alpha_ratio}$\alpha$ (a measure of band non-parabolicity) is plotted against the curvature effective mass $m_\text{c}$. The colour scale shows the ratio between the masses calculated using a Fermi--Dirac weighting and no weighting. We use the data from Table\ \ref{masstable}, plus data for two extra valence bands in \ce{Cu2ZnSnS4}, which, due to crystal field splitting, have a higher binding energy than the valence band maximum. All results are from data obtained using the HSE06 functional with spin-orbit coupling.} 
\end{figure}

We compare the weighted to unweighted calculation methods by considering the ratio of their effective mass values as a function of effective mass and non-parabolicity (Fig.\ \ref{mass_alpha_ratio}).
The weighted effective mass approach approximates the dispersion relation as parabolic, with a modified curvature due to flattening of the dispersion away from the band edge.
As a result, we calculate heavier effective masses using the weighted least-squares method compared to the unweighted least-squares method, which samples a smaller range of reciprocal space. 
For CdTe, however, the curvature effective mass calculated using the weighted least-squares approach is less than the value calculated using the unweighted approach. This is due to a non-parabolic flattening at the top of the valence band (Fig. S9), a feature which is seen only when spin-orbit effects are included. 
We find that the weighted---unweighted ratio is large for the heavy holes in CZTS and GaAs, for which there are a large number of eigenstates available at energies where the occupation probability, determined by the Fermi--Dirac distribution, is appreciable.
The ratio is also large for the MAPI valence band, where the flattening of the dispersion away from the band edge is appreciable due to a large $\alpha$ value.

The weighted least-squares fit presented here avoids polynomial fitting to an arbitrarily chosen energy range and is less sensitive to the sampling density used than the unweighted method, and is therefore an improvement. 
For high-curvature parabolic bands each of the three methods gives similar results.
For low-curvature bands the difference between the calculation methods is greatest.

\subsection{Sensitivity of the curvature effective mass to electronic structure method}

\begin{figure*}[tb] 
\includegraphics[width=\textwidth]{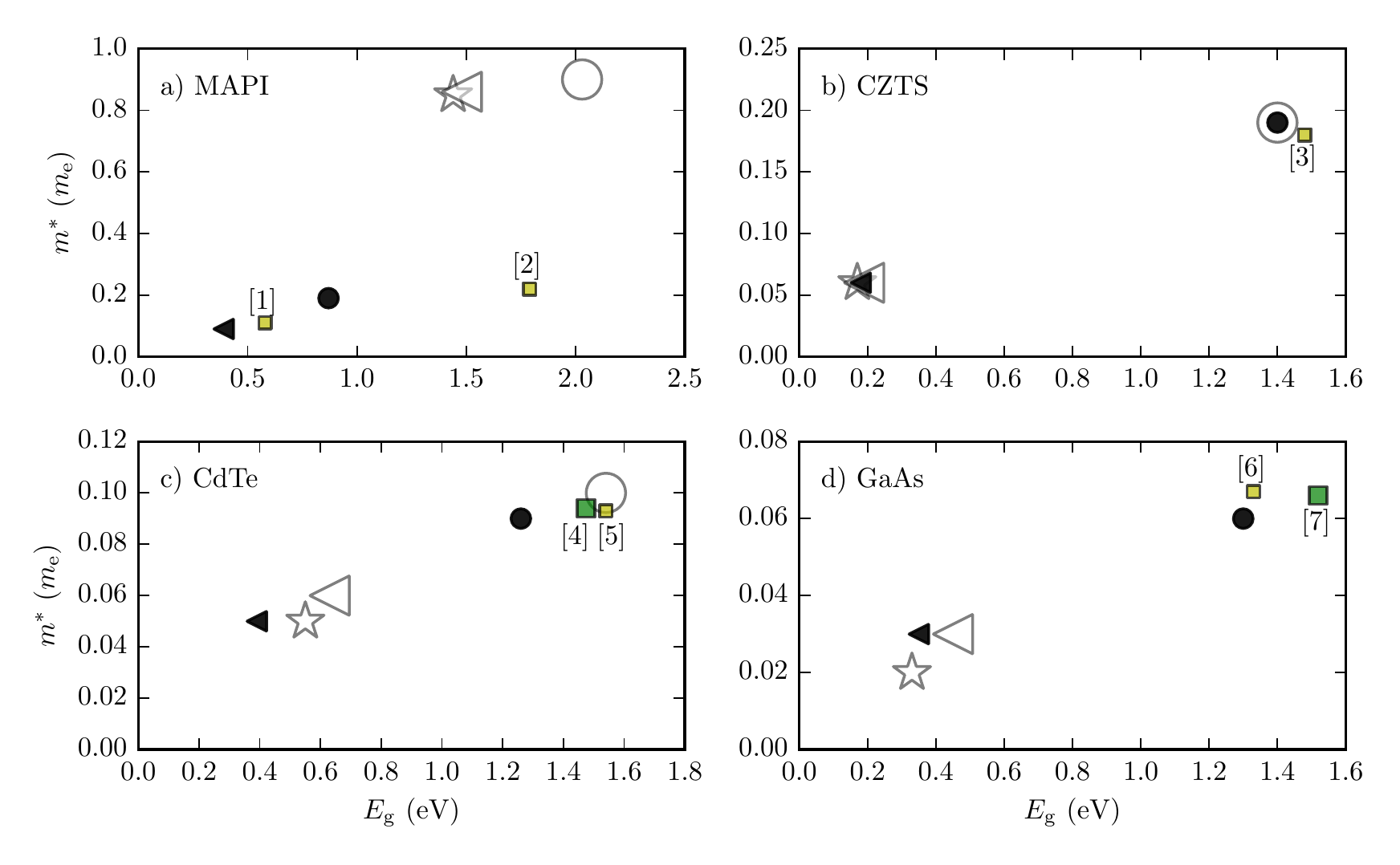}
\caption{\label{m*_bandgap_plot}The effective mass of conduction band electrons in the [100] direction is plotted against the band-gap. We calculate the effective mass using a weighted quadratic least-squares fit as outlined in the Methods section. Filled shapes denote a calculation performed with spin-orbit coupling. Stars, triangles and circles denote the use of the LDA, PBEsol and HSE06 exchange-correlation functionals respectively. Yellow squares denote results from other computational studies: [1] DFT+SoC\cite{Filip2015} [2] GW\cite{Filip2015} [3] DFT+SoC\cite{Liu2012} [5] GW\cite{Deguchi2016}  [6] 30-band $k\cdot p$ method.\cite{Richard2004} Green squares denote results from experimental studies [4] (obscured by [5]) cyclotron resonance\cite{Madelung2004} [7] cyclotron resonance.\cite{Madelung2004}}
\end{figure*}

In the previous section we considered how calculated values of the curvature effective mass depend on the details of the numerical approach chosen. 
In this section we examine how these calculated values vary with the choice of exchange-correlation functional and inclusion of spin-orbit coupling effects. 
Fig.\ \ref{m*_bandgap_plot} shows the weighted curvature effective mass against the band-gap, at different levels of theory. 
For CZTS, CdTe and GaAs we find that the local (LDA) and semilocal (GGA) approximations for the exchange energy underestimate the band-gap whereas the HSE06 hybrid functional gives values closer to those found experimentally. It is well established that HSE06 gives, in general, more accurate band-gaps than local and semilocal approximations.\cite{Deak2010,Heyd2003}

Tight-binding theory predicts that the effective mass is inversely proportional to the coupling strength.\cite{Kittel2005} 
An under prediction of band-gap means an over prediction of coupling strength and thus an under prediction of effective mass. 
This is observed in the positive correlation observed between the band-gap and the effective mass shown in Fig.\ \ref{m*_bandgap_plot}.
For CZTS, CdTe and GaAs, the effective mass calculated with the HSE06 functional agrees with experimental data to $0.01m_\mathrm{e}$.
The electron effective mass in CZTS is particularly sensitive to the exchange-correlation functional; the value calculated using the HSE06 functional ($0.19\,m_{\text{e}}$) is over three times that calculated using the LDA or GGA functional ($0.06\,m_\text{e}$).

MAPI has an experimental band-gap of $1.53\,\mathrm{eV}$,\cite{Liu2015} that is in good agreement with the value calculated using PBEsol with no spin-orbit coupling, due to a cancellation of errors. 
Whereas spin-orbit coupling has a negligible impact of the effective mass values calculated for CdTe, GaAs and CZTS,
it has a large influence on the effective mass values calculated for MAPI;
this is due to the large atomic charge of lead. 
Without spin-orbit coupling the MAPI effective mass values are signficantly overestimated. 

A reduced effective mass $((m^{*-1}_{\text{e}}+m^{*-1}_{\text{h}})^{-1})$ of $0.104\,m_{\text{e}}$ has been reported experimentally for MAPI.\cite{Miyata2015} 
This mass is extracted from magneto-absorption measurements where there is an interaction between the charge carrier and optical phonon modes.
This ``phonon drag" results in an experimentally measured mass which is heavier than the non-interacting mass extracted from a DFT band structure.
The experimental value is equal to the reduced mass calculated in the [100] direction using the hybrid HSE06 functional with spin-orbit coupling ($m^*_{\text{h}}=0.23$, $m^*_{\text{e}}=0.19$). 
Taking a harmonic average across the three directions in reciprocal space, we reach an average effective mass of $0.07\,m_{\text{e}}$, which is below that measured experimentally.

\subsection{Sensitivity of the Kane dispersion parameters to electronic structure method}

Until this point we have focused on the curvature effective mass which can be described with a single parameter.
This definition of effective mass is valid when the dispersion relation is well-approximated by a parabola.
For non-parabolic dispersions, a more accurate description is given by the Kane quasi-linear dispersion. 
In this formulation, the effective mass is described by two parameters: the transport effective mass at the band edge, $m_{\text{t},0}$, and the $\alpha$ parameter (Eqn.\ \ref{kanemass}).
We now focus on the sensitivity of the Kane dispersion parameters to the choice of exchange-correlation functional and to the inclusion of spin-orbit coupling effects. 

\begin{figure}[tb]
\includegraphics[width=\textwidth]{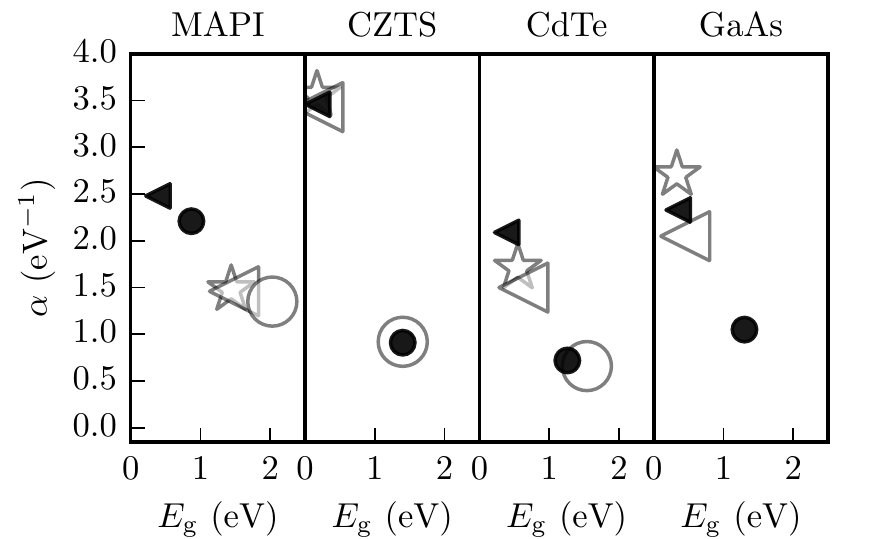} 
\caption{\label{alpha_bandgap_plot}$\alpha$ (a measure of band non-parabolicity) is plotted against band-gap at various levels of theory. $\alpha$ is calculated for the conduction band in the [100] direction. We use filled shapes to denote a calculation with spin-orbit coupling. We use stars, triangles and circles to denote the use of the LDA, PBEsol and HSE06 exchange-correlation functionals, respectively. }
\end{figure}

For all the materials considered here, the parameter $\alpha$ is inversely correlated with the band-gap of the material (Fig.\ \ref{alpha_bandgap_plot}). 
In CZTS, for example, the conduction band is highly non-parabolic ($\alpha\approx3.5\,\mathrm{eV}^{-1}$) at lower levels of theory (LDA/GGA).
When a hybrid functional is used the band-gap increases and the conduction band becomes more parabolic ($\alpha\approx1\,\mathrm{eV}^{-1}$).
We attribute this to the $k\cdot p$ interaction between the conduction and valence bands,\cite{Kane1957} which makes non-parabolicity particularly pronounced in narrow band-gap semiconductors such as GaAs.\cite{Szmyd1990}
Local and semi-local density functional approximations, which underestimate the band-gap, lead to an enhanced $k\cdot p$ interaction and an overestimated non-parabolicity.

\begin{figure}[tb]
\includegraphics[width=\textwidth]{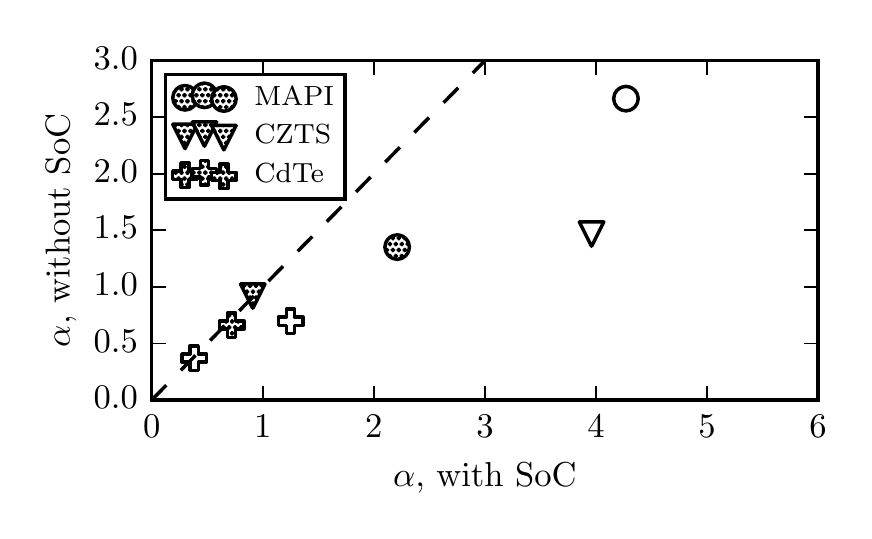} 
\caption{\label{alpha_SoC}$\alpha$ (a measure of band non-parabolicity, eV$^{-1}$) in the [100] direction, calculated with and without spin-orbit coupling. We use hatched shapes to denote an electron in the conduction band, and empty shapes to denote a hole in the valence band. Note the scale; the dashed line indicates where the values would lie if spin-orbit coupling had no influence on the value of $\alpha$. The hybrid HSE06 exchange-correlation functional was used for all calculations. }
\end{figure}

The non-parabolicity of bands in CZTS\cite{Ito2015} and MAPI\cite{Brivio2014,Mosconi2017} has been previously attributed to the spin-orbit interaction. 
For the CZTS conduction band we find that the amount of non-parabolicity is determined by the band-gap and so, indirectly, by the exchange-correlation functional used. Spin-orbit coupling has a small effect (Fig.\ \ref{alpha_SoC}). 
Spin-orbit coupling has a larger effect on the non-parabolicity of the CZTS valence band, which has an $\alpha$ value of $3.96\,\mathrm{eV}^{-1}$ when calculated with spin-orbit coupling and $1.47\,\mathrm{eV}^{-1}$ when calculated without. 
For MAPI, both the valence and conduction bands are more non-parabolic when spin-orbit coupling is included.

Our results show that it is important to reproduce experimental band-gaps to accurately calculate effective mass parameters.
The effective mass is under-estimated and band non-parabolicity is over-estimated using local and semi-local density functional approximations.
This illustrates the importance of using a method that gives an accurate band gap predictions when calculating effective mass parameters.
For MAPI, to calculate accurate effective mass values and capture the full extent of band non-parabolicity, we must also account for spin-orbit effects.

\subsection{Consequences of non-parabolicity on the optical and transport properties of MAPI}

The effective mass is commonly calculated and compared across materials because it can be used to model differences in various optical and transport properties.
The many relationships between effective mass and various experimental observables also means that the effective mass can be inferred from a number of different measurements.
In practice, effective masses derived from different experimental properties can have inconsistent values.
For materials with highly non-parabolic bands, experimental observables, and the corresponding derived effective mass values, often vary with temperature and carrier concentration.

In interpreting the photophysical behaviour of MAPI a range of effective mass values can be found in the literature, from $0.09\,m_{\text{e}}$--$0.38\,m_{\text{e}}$.\cite{Miyata2015,Tanaka2003,Hirasawa1994,Yang2015,Manser2014,Price2015}
Here we focus on the results from transient absorption spectroscopy (TAS) as an illustrative dataset.\cite{Manser2014,Yang2015,Price2015}
Manser \textit{et al}.\ calculated the Burstein--Moss shift in MAPI up to a carrier concentration of ${1.5 \times 10^{19}\,\mathrm{cm}^{-3}}$.\cite{Manser2014} By assuming a parabolic band dispersion, they calculated a reduced effective mass at the band edge of ${0.3\,m_{\text{e}}}$, which is larger than effective mass values calculated from magneto-absorption experiments\cite{Miyata2015,Tanaka2003,Hirasawa1994} and theory.\cite{Brivio2014,Umari2014} 
They suggest that this discrepancy could be due to `band-gap renormalisation', a concept that incorporates many physical phenomena, including band non-parabolicity.\cite{Walsh2008}
In a separate study which included band-gap renormalisation, Yang \textit{et al}.\ reported that TAS spectra, up to a carrier concentration of ${5.5 \times 10^{18}\,\mathrm{cm}^{-3}}$, can be modelled with an electron effective mass value of ${0.23\,m_{\text{e}}}$.\cite{Yang2015} Assuming that the electron and hole effective masses are equal, we use this value to calculate a reduced effective mass value of ${0.12\,m_{\text{e}}}$.
A third estimate of the reduced effective mass is given by Price \textit{et al}.\ who have developed a model that includes photoinduced changes to the refractive index.\cite{Price2015} 
This approach, which assumes a parabolic dispersion, and fits to TAS data up to a carrier concentration of ${6.4 \times 10^{18}\,\mathrm{cm}^{-3}}$, gives a reduced effective mass of $0.14\,m_{\text{e}}$.
The latter two approaches give reduced effective mass values in line with results from magneto-absorption experiments\cite{Miyata2015,Tanaka2003,Hirasawa1994} and previous theory.\cite{Brivio2014,Umari2014} 

In this section we will quantify the extent to which non-parabolicity affects the Burstein--Moss band-gap shift, the optical effective mass, and the polaron mobility in MAPI.
To do so, we first assume a Kane quasi-linear dispersion and then predict how these observables will vary as a function of carrier concentration.

\subsubsection{Burstein--Moss band-gap shift}

One consequence of the Pauli exclusion principle is that an increase in carrier concentration can push the Fermi level into the conduction, or valence, band.
This band filling is observed as an increase in the optical band-gap, called a ``Burstein--Moss shift''.
For a parabolic dispersion, the magnitude of the Burstein--Moss shift is given by\cite{Moss1954}
\begin{equation} \label{parabolic_shift}
\Delta_{\text{BM}} =\frac{\hbar^2}{2m^*}(3\pi^2n_{\text{e}})^{\frac{2}{3}},
\end{equation}
where $m^*$ is the reduced effective mass and $(3\pi^2n_{\text{e}})^{2/3}$ is the Fermi wavevector up to which all states are occupied under the free electron model. 
The Burstein--Moss shift is most commonly considered in the context of degenerately doped semiconductors. Here we consider photo-excited carriers in an undoped material (MAPI).

Above a critical carrier concentration the electron distribution becomes degenerate and it is in this regime that the Burstein--Moss shift occurs.
Manser \textit{et al}.\ have reported an abrupt onset of the Burstein--Moss effect at carrier concentrations of $7.5\times10^{17}\,\mathrm{cm}^{-3}$.\cite{Manser2014}
They attribute this behaviour to trap filling, with reference to an estimated trap density of $2\times10^{17}\,\mathrm{cm}^{-3}$.\cite{Xing2014}
We add to this trap density the density at which the electron polaron wavefunctions overlap, $4\times10^{17}\,\mathrm{cm}^{-3}$,\cite{Frost2017} as given by the phenomonological Mott criterion.\cite{Mott1949}
This gives a total of $6\times10^{17}\,\mathrm{cm}^{-3}$, which is closer to the critical carrier concentration reported by Manser \textit{et al}.
This exceeds the predicted maximum carrier concentration of $\sim10^{16}\,\mathrm{cm}^{-3}$ under AM1.5 solar illumination.\cite{Herz2016} 
Carrier concentrations of up to $10^{18}\,\mathrm{cm}^{-3}$ have been reported for concentrator systems,\cite{Law2014} which would correspond to the degenerate regime.
Furthermore, carrier concentrations of up to $10^{19}\,\mathrm{cm}^{-3}$ are achievable under laser excitation, as in photoluminescence and transient absorption spectroscopy measurements,\cite{Richter2016}
and there has been recent research interest in developing a hybrid halide perovskite lasing material which would require carrier concentrations of up to $10^{21}\,\mathrm{cm}^{-3}$.\cite{Gao2017}

\begin{figure}[tb]
\includegraphics[width=\textwidth]{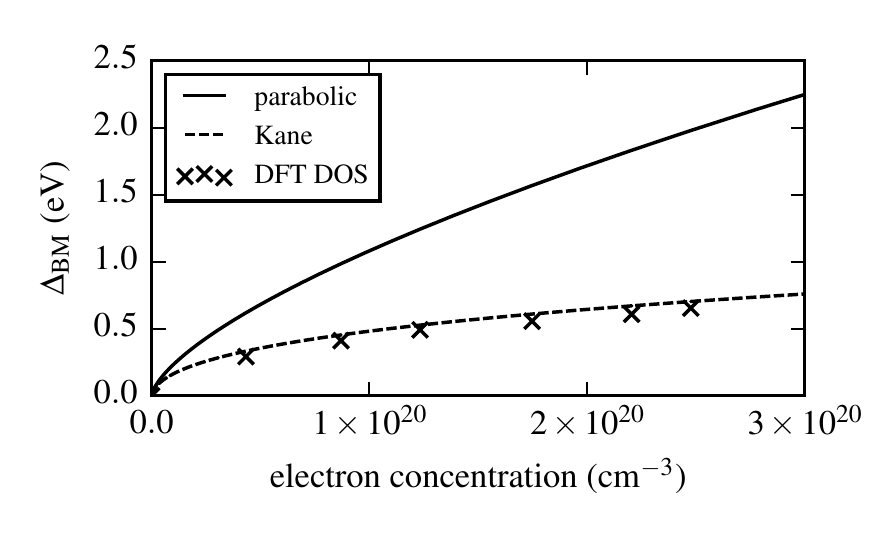}
\caption{\label{burstein_moss_plot} The Burstein--Moss band-gap shift is plotted as a function of carrier concentration. The shift is calculated assuming a parabolic dispersion, Kane quasi-linear dispersion and from a DFT calculated density-of-states. Beyond this carrier concentration range we begin to fill bands higher in energy (the secondary band in MAPI is calculated to be at $0.78\,\mathrm{eV}$ above the conduction band minimum). To average over directions in reciprocal space we use the geometric mean of the effective mass and the arithmetic mean of $\alpha$.}
\end{figure}

Fig.\ \ref{burstein_moss_plot} shows the predicted Burstein--Moss shift for MAPI, calculated using Eqn.\ \ref{parabolic_shift}, with $m^*$ obtained first from parabolic fits to the band dispersion and second from the energy-dependent mass given by Eqn.\ \ref{kanemass}.
We compare our results to the Burstein-Moss shift calculated directly from the DFT density-of-states. For the latter calculation we make no assumption about the form of the dispersion; the shift is defined as equal to the highest energy level filled at a given concentration of electrons which have thermalised to the band edge.
Using the parabolic approximation leads to a large overestimation of the Burstein--Moss shift (Fig.\ \ref{burstein_moss_plot}). 
Walsh \textit{et al}.\cite{Walsh2008}\ demonstrated that non-parabolicity can cause this discrepancy, as
band flattening corresponds to an increased density-of-states, which results in the Fermi level increasing more slowly than in the parabolic case. 
Here, we account for non-parabolicity by substituting the energy-dependent mass given by Eqn.\ \ref{kanemass} into Eqn.\ \ref{parabolic_shift} and solving self-consistently.
With this amendment we obtain a good agreement with the density-of-states data. This demonstrates that the Kane quasi-linear dispersion provides a suitable approximation to the electronic dispersion relation at high carrier concentrations.
 
\subsubsection{Optical effective mass}

To calculate the effective mass measured in optical experiments, we integrate the analytic expression given in Eqn.\ \ref{opt}, with $E(k)$ given by the Kane dispersion in Eqn.\ \ref{kane}. 
We set the Fermi level equal to the Burstein--Moss shift calculated using the self-consistency procedure described above, which introduces a dependance on carrier concentration. 

\begin{figure}[tb]
\includegraphics[width=\textwidth]{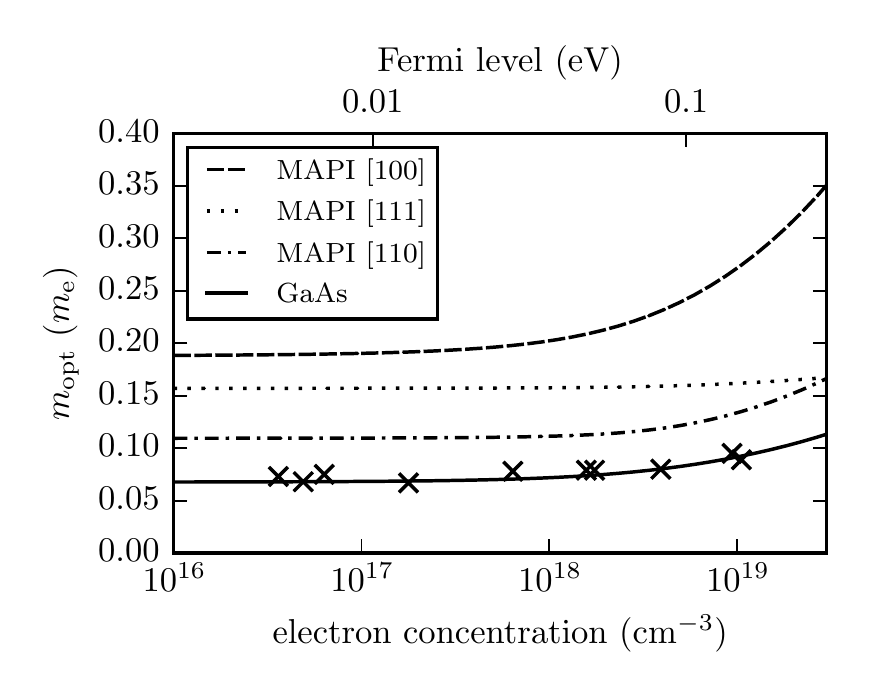}
\caption{\label{optical_concentration_plot} The conduction band electron optical effective mass is plotted as a function of carrier concentration. Values for \ce{CH3NH3PbI3} in the [100] (dash line), [111] (dot line) and [110] (dash-dot line) directions are given. To validate our results we include results for GaAs (continuous line) and compare this against experimental data (crosses).\cite{Raymond1979} The Kane quasi-linear dispersion in the [100] direction does not give a good approximation to the DFT-calculated dispersion beyond carrier concentrations of $3\times10^{19}\,\mathrm{cm}^{-3}$. }
\end{figure}

There is a significant increase in the optical effective mass in the [100] direction, from $0.19\,m_\text{e}$ at the band edge to $0.35\,m_\text{e}$ at a carrier concentration of $3\times10^{19}\,\mathrm{cm}^{-3}$ (Fig.\ \ref{optical_concentration_plot}). 
To put this in context, the effective mass values calculated from a parabolic fitting to GW+SoC and PBE+SoC band structures differ by up to only $0.03\,m_\text{e}$.\cite{Umari2014}
At high carrier concentrations, band filling has a much larger effect on the optical effective mass than differences in the level of theory used to calculate the band structure.
This observation illustrates the importance of accounting for non-parabolic dispersion for accuracte calculations of optical effective mass.

\begin{table*}[tb]\centering
\caption{\label{masstable} The polaron mobility for a conduction band electron in \ce{CH3NH3PbI3}, using the optical effective mass calculated in the [110] direction.}
\begin{tabular}{@{}ccccccr@{}}\toprule
concentration (cm$^{-3}$) && effective mass $m_\text{opt}$ ($m_\text{e}$) && mobility ($\text{cm}^2\text{V}^{-1}\text{s}^{-1}$) && application \\ \colrule
10$^{15}$ && 0.11 && 158 && solar cell, standard operating conditions \\
10$^{18}$ && 0.11 && 158 && concentrator system \\
10$^{19}$ && 0.13 && 120 && photoluminescence\\
10$^{20}$ && 0.23 && 46 && lasing material \\
\botrule
\end{tabular}
\end{table*}

\subsubsection{Polaron mobility}

The Drude model describes charge-carrier motion as a classical gas, propagating freely before scattering elastically after a fixed relaxation time. 
For a particle with charge $q$, the mobility $\mu$, relaxation time $\tau$, and effective mass $m^*$ are related via

\begin{equation}
\mu = \frac{q\tau}{m^*}.
\end{equation}

In this model, mobility is inversely proportional to the effective mass. 
However for most scattering processes, the relaxation time is itself a function of the effective mass. 
In strongly scattering systems, there is no free propagation, and scattering cannot be reduced down to discrete events. 
Here we study the effects of band non-parabolicity by calculating electron polaron mobility as a function of the optical effective mass. 

MAPI is polar and soft which leads to a large dielectric electron-phonon coupling.
Scattering from polar optical phonon modes is the process that limits the charge carrier mobility in MAPI at room temperature.\cite{Wright2016} 
We have calculated the mobility of a large polaron at $T=300\,\mathrm{K}$ using a parameter-free variational method based on a Feynman path integral approach.\cite{frost2017calculating}
This solves the quantum-field problem of an individual electron interacting by the dielectric electron-phonon coupling with an infinite field of polar phonon modes. 
The low-field mobility is recovered from a contour integration of the polaron response function. 
The input parameters for this model are the effective mass, the high- and low-frequency dielectric constants and a characteristic phonon mode, and we have used values from previous work.\cite{frost2017calculating}

\begin{figure}[tb]
\includegraphics[width=\textwidth]{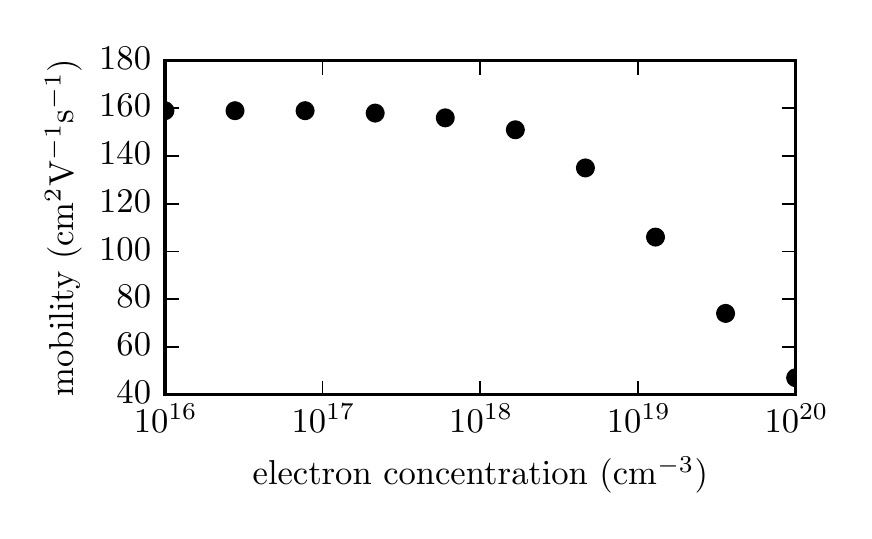}
\caption{\label{mobility_plot} Polaron mobility, limited by optic-mode scattering, is plotted as a function of carrier concentration for a conduction band electron in the [110] direction of \ce{CH3NH3PbI3}. This is calculated using parameters from previous work:\cite{frost2017calculating} optical dielectric constant $=4.5$, low-frequency dieletric constant $=24.1$ and characterisitic phonon frequency $=2.25\times10^{12}\,\mathrm{THz}$.}
\end{figure}

At carrier concentrations up to $10^{18}\,\mathrm{cm}^{-3}$, and using the optical effective mass in the [110] direction as an example, the mobility for a conduction band electron in MAPI is calculated to be $158\,\mathrm{cm^2V^{-1}s^{-1}}$. 
At higher carrier concentrations, our calculated mobility decreases, falling to $\sim30\%$ of the low carrier concentration value by $10^{20}\,\mathrm{cm}^{-3}$ (Fig.\ \ref{mobility_plot} and Table \ref{masstable}).
It is difficult to compare our results to experiment as there are strong variations in the experimental values reported, from $2.5\,\mathrm{cm^2V^{-1}s^{-1}}$ to $600\,\mathrm{cm^2V^{-1}s^{-1}}$. The average value taken over multiple results for a single crystal is $73 \pm 58\,\mathrm{cm^2V^{-1}s^{-1}}$.\cite{Herz2017} 
The upper end of this range is comparable to, but below, our predicted mobility at low carrier concentration. 
This is to be expected; in our model for polaron mobility we only consider the harmonic dielectric response, whilst experimental results will include all anharmonicity, which will increase the dielectric response and decrease carrier mobility.

The polaron model used is formally for an individual electron. 
We use this model to understand how the changes in effective mass due to doping (electron correlation) affect the mobility. 
A previous theoretical analysis\cite{Lemmens1977} suggests polarons are non-interacting (as shown by ground state energy) at densities below half the Mott critereon. 
From our previous work\cite{Frost2017} this would be $2\,\times\,10^{17}\,\mathrm{cm}^{-3}$. 
This threshold, where the polarons start to interact, corresponds to the onset of electron---electron scattering. 
Even if this scattering contribution is neglected, however, we still predict a strong reduction in mobility. 
Our result agrees with a previous report which states that mobility is particularly sensitive to effective mass.\cite{Ponce2018} 

Our results predict that, in MAPI, non-parabolicty has a negligible effect at carrier concentrations up to $10^{18}\,\mathrm{cm}^{-3}$.
At a concentration of $4\times10^{18}\,\mathrm{cm}^{-3}$, the Burstein--Moss band-gap shifts predicted using the parabolic and Kane quasi-linear dispersion relations differ by only $0.026\,\text{eV}$.
This is equal to the typical energetic disorder at room temperature ($k_{\mathrm{B}}T=0.026\,\mathrm{eV}$), and so we predict the parabolic approximation to be reasonable up to this carrier concentration.
At carrier concentrations above $4\times10^{18}\,\mathrm{cm}^{-3}$, band non-parabolicity becomes more significant, and must be accounted for to accurately predict the Burstein-Moss band-gap shift, optical effective mass and polaron mobility (Table\ \ref{masstable}). 
Due to non-parabolic dispersion, at a concentration of $10^{20}\,\mathrm{cm}^{-3}$ the optical effective mass for a conduction band electron in the [110] direction of MAPI increases by a factor of two relative to the low carrier-concentration value, and the polaron mobility decreases by a factor of three. 

\section{Summary and Conclusions}

The effective mass plays a foundational role in the quantitative description of device properties, and is a key parameter when assessing the potential of new functional materials. 
Several alternative definitions exist for the effective mass, and each of these values calculated from electronic structure data also depends on the details of the chosen numerical method.

In this paper, we have introduced a thermally weighted least-squares method to calculate the curvature effective mass in the parabolic approximation. This method is physically intuitive, and by regularising the least squares fit with weights from a Fermi--Dirac distribution, it accounts for thermal sampling of relevant states by charge carriers. 
We have shown that this method is more robust to sampling density than the popular unweighted least-squares or finite-difference methods to the sampling density in reciprocal space.

We then moved beyond the parabolic approximation to consider band structures described by Kane quasi-linear dispersion.
The $\alpha$ parameter provides a good description of the dispersion away from band edge, and could be used as a screening parameter for materials to operate at high carrier concentrations; for example transparent conducting oxides.
For the four materials in this study, \ce{CdTe}, \ce{GaAs}, \ce{MAPI} and \ce{CZTS}, conventional semi-local exchange-correlation functionals underestimate band gaps, and overestimate the degree of non-parabolicity. More accurate hybrid functionals are required to obtain accurate effective mass values.
We have also shown the importance of including spin-orbit coupling effects for describing the non-parabolicity of the valence bands via the $\alpha$ parameter.

Finally, we have considered how non-parabolicity affects predicted optical and transport properties at high carrier concentrations,
for the lead halide perovskite MAPI, which is the most highly non-parabolic of the four materials considered here.
Using effective masses calculated in the parabolic approximation significantly overestimates the Burstein--Moss shift, 
whilst using effective masses calculated from the Kane dispersion gives a Burstein--Moss shift in good agreement with the shift calculated directly from the density-of-states.
By using the Burstein--Moss shift as a proxy for the Fermi level, we calculated the change in electron effective mass and mobility as a function of carrier concentration. 
The assumption previously used in the literature---that the effective mass of MAPI is independent of carrier concentration---is not valid.
At carrier concentrations above $10^{18}\,\mathrm{cm}^{-3}$, non-parabolicity must be built into photophysical models to give accurate values for the effective mass and derived properties.
We expect similar adjustments should be made to the predicted optical and transport properties of other semiconductors where non-parabolicity is significant at relevant carrier concentrations.

\acknowledgments

This work used the ARCHER UK National Supercomputing Service (http://www.archer.ac.uk) which we have access to via our membership of the UK's HEC Materials Chemistry Consortium (funded by EPSRC Grant No. EP/L000202). We are also grateful to the UK Materials and Molecular Modelling Hub for computational resources, which is partially funded by EPSRC Grant No. EP/P020194/1. L.~W. and J.~M.~F. are funded by the EPSRC (Grant Numbers EP/L01551X/1 and EP/R005230/1 respectively). A.~W. and B.~J.~M. acknowledge support from the Royal Society (Grant Numbers UF150657 and UF130329 respectively).
\\
 
\textbf{Data Access Statement} \\ 

Input crystal structures, calculated electronic structure data, and a Jupyter notebook outlining the key calculation steps used in this paper are available in two online repositories,\cite{Whalley2018,Whalley2018b} which have been published alongside the \textsc{effmass} package for calculating effective masses.\cite{Whalley2018a}
Polaron mobilities were calculated with the \textsc{PolaronMobility.jl} package.\cite{frost2017calculating,Frost2018,PolaronMobilityGithub}
All data and packages are available under the MIT license.

\end{document}